\documentclass[12pt,a4paper]{article}
\usepackage{graphicx}
\usepackage[T1]{fontenc}
\usepackage[utf8]{inputenc}
\usepackage{textcomp}
\usepackage[sc,osf]{mathpazo}
\usepackage{a4wide}  
\usepackage{latexsym,amsthm,amsfonts,amsmath,mathrsfs,amssymb}
\usepackage{dsfont}
\usepackage{accents}
\usepackage[nosort]{cite}
\usepackage{booktabs} 
\usepackage[unicode,implicit]{hyperref}
\hypersetup{%
  pdftitle    = {On the generalized Komar charge of Kaluza-Klein theories and
    higher-form symmetries}
  pdfkeywords = {gravity,  symmetries, conserved charges, Komar charge,
    Komar intergral, Kaluza-Klein, black holes},
  pdfauthor   = {Gabriele Barbagallo, Jos\'e Luis V. Cerdeira, Carmen G\'omez-Fayr\'en, Patrick Meessen and Tom\'as Ort\'{\i}n},
  plainpages  = true,
  colorlinks  = true,
  citecolor   = blue,
  urlcolor    = red,
  linkcolor   = black
}
\newcommand{\hepth}[1]{{\tt
\href{http://www.arXiv.org/abs/hep-th/#1}{hep-th/#1}}}
\newcommand{\grqc}[1]{{\tt
\href{http://www.arXiv.org/abs/gr-qc/#1}{gr-qc/#1}}}

\newcommand{\arxiv}[1]{{\tt arXiv:\href{http://www.arXiv.org/abs/#1}{#1}}}
\newcommand{\doi}[1]{{\tt DOI:\href{http://dx.doi.org/#1}{#1}}}

\allowdisplaybreaks

\makeatletter
\@addtoreset{equation}{section}
\makeatother

\begin{document}

\begin{flushright}
\small
IFT-UAM/CSIC-25-063\\
June 16\textsuperscript{th}, 2025\\
\normalsize
\end{flushright}

\vspace{.2cm}

\begin{center}

  {\Large {\bf On the generalized Komar charge\\[.5cm] of Kaluza--Klein
      theories and higher-form symmetries}}
 
\vspace{1cm}

\renewcommand{\thefootnote}{\alph{footnote}}

{\sl Gabriele Barbagallo,$^{1,}$}\footnote{Email: {\tt gabriele.barbagallo[at]estudiante.uam.es}}
{\sl Jos\'e Luis V.~Cerdeira}$^{2,}$\footnote{Email: {\tt jose.verez-fraguela[at]estudiante.uam.es}}
{\sl Carmen G\'omez-Fayr\'en,$^{1,}$}\footnote{Email: {\tt carmen.gomez-fayren[at]estudiante.uam.es}}\\[.5cm]
{\sl Patrick Meessen$^{3,4}$}\footnote{Email: {\tt meessenpatrick[at]uniovi.es}}
{\sl and Tom\'{a}s Ort\'{\i}n}$^{1,}$\footnote{Email: {\tt Tomas.Ortin[at]csic.es}}

\setcounter{footnote}{0}
\renewcommand{\thefootnote}{\arabic{footnote}}
\vspace{1cm}

${}^{1}${\it\small Instituto de F\'{\i}sica Te\'orica UAM/CSIC\\
C/ Nicol\'as Cabrera, 13--15,  C.U.~Cantoblanco, E-28049 Madrid, Spain}

\vspace{0.2cm}

${}^{2}${\it\small Instituto de F\'{\i}sica Corpuscular (IFIC), University of
  Valencia-CSIC,\\
Parc Cient\'{\i}fic UV, C/ Catedr\'atico Jos\'e Beltr\'an 2, E-46980 Paterna, Spain}

\vspace{0.2cm}

${}^{3}${\it\small HEP Theory Group, Departamento de F\'{\i}sica, Universidad de Oviedo\\
  Calle Leopoldo Calvo Sotelo 18, E-33007 Oviedo, Spain}\\

\vspace{0.2cm}

${}^{4}${\it\small Instituto Universitario de Ciencias y Tecnolog\'{\i}as Espaciales
  de Asturias (ICTEA)\\ Calle de la Independencia, 13, E-33004 Oviedo, Spain}

\vspace{1cm}

{\bf Abstract}
\end{center}
\begin{quotation}
  {\small The generalized Komar $(d-2)$-form charge can be modified by the
    addition of any other on-shell closed (\textit{conserved}) $(d-2)$-form
    charge. We show that, with Kaluza--Klein boundary conditions, one has to
    add a charge related to the higher-form symmetry identified in
    Ref.~\cite{Gomez-Fayren:2024cpl} to the naive Komar charge of pure
    5-dimensional gravity to obtain a conserved charge charge whose integral
    at spatial infinity gives the mass. The new term also contains the
    contribution of the Kaluza--Klein monopole charge leading to
    electric-magnetic duality invariance.  }
\end{quotation}

\newpage
\pagestyle{plain}



\section{Introduction}

The impact and influence of Kaluza's discovery that some components of the
5-dimensional metric behaved as the 4-dimensional metric and Maxwell fields
\cite{Kaluza:1921tu} complemented with Klein's realization that, if the
5\textsuperscript{th} dimension is compact, the 4-dimensional charge and mass
of a massless and uncharged particle moving in 5 dimensions are proportional
to its momentum in that compact direction, inversely proportional to its size
and quantized \cite{Klein:1926tv} in the subsequent development of Theoretical
Physics cannot be overstated.\footnote{See Ref.~\cite{Appelquist:1987nr} for a
  review.}  The main difference (and a very important one) between the gauge
theories on which the Standard Model is based and Kaluza-Klein theories is
that the extra dimensions of the former are not spacetime dimensions along
which particles and gravity can propagate as in the second. As a matter of
fact, the first non-Abelian Yang--Mills type theory was constructed by Pauli
using Kaluza--Klein (KK) compactification in a space with SU$(2)$ symmetry
\cite{Straumann:2000zc} and the use of this mechanism in Supergravity and
Superstring theories has played a fundamental role in the search for a unified
theory of all interactions \cite{Duff:1986hr,Duff:2025tot,Ibanez:2012zz}.

The KK paradigm ``Physics in the uncompactified (\textit{lower}) dimensions is
just a manifestation of Physics in the total spacetime manifold
(\textit{higher dimensions})'' means that we should be able to derive no
matter what lower-dimensional results working directly in higher
dimensions. In this article we are concerned with Noether--Wald and
generalized Komar charges and their use to study the thermodynamics of stringy
black holes, finding their Smarr formulas and deriving the first law with all
their charges and chemical potentials. These black holes are classical
solutions of lower-dimensional Supergravity theories most of which can be
derived by KK dimensional reduction from an 11- or 10-dimensional
Supergravity\footnote{For a recent review with many references see, for
  instance, Ref.~\cite{Ortin:2024slu}.}  and it follows that we should be able
to derive those Smarr formulas and first laws directly in 10 or 11 dimensions.

In order to achieve this goal 
\begin{enumerate}
\item We need a higher-dimensional interpretation of all the conserved
  charges carried by the lower-dimensional black-hole solutions. This requires
  a complete dictionary between the local and global symmetries of the higher-
  and lower-dimensional theories.
\item We need a higher-dimensional interpretation of the associated
  chemical potentials.
\item We need a good understanding of the relations between the black-hole's
  higher- and lower-dimensional geometries. In particular, we need to know the
  relation between their event or Killing horizons.
\item With those interpretations we should be able to derive the Smarr
  formulas and first laws directly from the Noether--Wald and generalized
  Komar $(d-2)$-form charges ($d=10,11$).
\end{enumerate}

In Refs.~\cite{Gomez-Fayren:2023wxk,Gomez-Fayren:2024cpl} we have made
progress in some of the points of this program using pure 5-dimensional
Einstein gravity compactified on a circle. In Ref.~\cite{Gomez-Fayren:2023wxk}
we studied the 5-dimensional geometry of the Killing horizons of static
electrically-charged 4-dimensional black holes, finding that they are only
stationary due to the uplifting to 5 dimensions of the timelike 4-dimensional
Killing vector, which is a linear combination of the latter with the Killing
vector generating translations along the compact direction. We also found that
there is frame-dragging along the compact direction and that the velocity of
the horizon in that direction is the 4-dimensional electrostatic potential. We
also studied how the 5-dimensional Noether--Wald and Komar charges associated
to the uplifted timelike Killing vector give rise to the 4-dimensional Smarr
formula and 1\textsuperscript{st} law. The additional terms proportional to
the the Killing vector that generates translations along the compact direction
give contributions proportional to the momentum in that direction which, as
Klein found out, is seen as electric charge in 4 dimensions. However, we could
not find the term proportional to the 4-dimensional magnetic charge in the
5-dimensional 1\textsuperscript{st} law and the related term in the
4-dimensional Smarr formula only appeared after performing a rather
\textit{ad hoc} trick. 

On the other hand, in Ref.~\cite{Gomez-Fayren:2024cpl} we found that the
origin of the global symmetry of the 4-dimensional theory is a
higher-form-type global symmetry which is only present in the 5-dimensional
theory with the topologically non-trivial KK boundary conditions (one compact
spatial dimension).

In this article we are going to show the relation between the higher-form
symmetry found in Ref.~\cite{Gomez-Fayren:2024cpl} and the trick that
allowed us to recover the full electric-magnetic duality-invariant Smarr
formula in 4 dimensions in Ref.~\cite{Gomez-Fayren:2023wxk}. We are going to
show that this trick is actually necessary if we want to have a generalized
Komar charge for the Killing vector that generates time translations whose
integral at infinity gives the mass.\footnote{It is important to clarify from
  the onset that in this context (5-dimensional gravity with KK boundary
  conditions) by mass of a solution we mean the mass of the 4-dimensional
  (dimensionally-reduced) solution in the 4-dimensional Einstein frame. While
  there are intrinsically 5-dimensional definitions of the mass in this context
  \cite{Bombelli:1986sb,Deser:1988fc}, we believe that the above is the most
  natural notion.}

It is not difficult to see why the 5-dimensional Komar charge does not give
the mass: the 4-dimensional metric and the 4-dimensional components of the
5-dimensional metric are related by a conformal transformation with conformal
factor $k_{\infty}/k$ where $k$ is the KK scalar. The Komar integral at
infinity simply picks the coefficient of $1/r$ in the the expansion of the
$tt$-component of the metric

\begin{equation}
g_{tt} \sim 1-\frac{2M}{r}\,.  
\end{equation}

\noindent
However, if, at infinity

\begin{equation}
k\sim k_{\infty}\left(1+\frac{\Sigma}{r}\right)\,,  
\end{equation}

\noindent
where $\Sigma$ is the scalar charge, the expansion of the $tt$-component of
the  5-dimensional metric will be

\begin{equation}
\label{eq:scalarchargeplusmass}
(k_{\infty}/k)g_{tt} \sim 1-\frac{2M+\Sigma}{r}\,.
\end{equation}

\noindent
The naive Komar integral will give a combination of the mass and the scalar
charge and we have to modify it in a consistent way if we want it to give just
the mass $M$.

As we are going to see, the trick used in Ref.~\cite{Gomez-Fayren:2023wxk}
amounts to the addition of an on-shell closed (\textit{conserved})
$(d-2)$-form charge to the naive generalized Komar charge (which is also
on-shell closed, by construction) as explained in Ref.~\cite{kn:CO}.  This
addition does not change the value of the Smarr formula, since one adds and
subtracts the same quantity,\footnote{The Smarr formula can be obtained by
  integrating the generalized Komar charge over spatial infinity and over the
  horizon and equating the two results
  \cite{Bardeen:1973gs,Carter:1973rla,Magnon:1985sc,Bazanski:1990qd,Kastor:2010gq}. The
  on-shell closedness of the generalized Komar charge and the fact that these
  two $(d-2)$-dimensional surfaces are the only two pieces of the boundary of
  some hypersurface imply this equality.} but, when we only use the
generalized Komar charge to compute the mass or other gravitational conserved
charges integrating at infinity, the additional term does contribute to the
result in a non-trivial way. In our case we will use this freedom to eliminate
unwanted contributions.

We are going to start by reviewing the setup we are going to work with (pure
5-dimensional gravity with Kaluza--Klein boundary conditions) and the results
of Refs.~\cite{Gomez-Fayren:2023wxk,Gomez-Fayren:2024cpl} that we are going to
use.

\section{The setup}

In this paper we are going to use the conventions of
Ref.~\cite{Ortin:2015hya}, with a mostly minus signature, the Vielbein
$e^{a}=e^{a}{}_{\mu}dx^{\mu}$ as gravitational field and differential-form
notation.\footnote{The Lorentz (spin) connection
  $\omega^{ab}=\omega_{\mu}{}^{ab}dx^{\mu}=-\omega^{ba}$ is assume to be the
  torsionless Levi-Civita connection $\omega^{ab}=\omega^{ab}(e)$
  \begin{equation}
    \mathcal{D}e^{a}
    \equiv de^{a}-\omega^{a}{}_{b}\wedge e^{b}
    =
    0\,,
  \end{equation}
  and its curvature is defined as
  \begin{equation}
    R^{ab}(\omega)
    \equiv
    d\omega^{ab} -\omega^{a}{}_{c}\wedge \omega^{cb}\,.
  \end{equation}
} Furthermore, we use hats to denote the 5-dimensional quantities.
In this language, the 5-dimensional Einstein--Hilbert action has the form

\begin{equation}
  \label{eq:5dEHactionVielbein}
  S[\hat{e}]
  =
  \frac{1}{16\pi G_{N}^{(5)}}
  \int
\tfrac{1}{3!}
      \hat{\varepsilon}_{\hat{c}\hat{d}\hat{e}\hat{a}\hat{b}}
      \hat{e}^{\hat{c}}\wedge \hat{e}^{\hat{d}} \wedge \hat{e}^{\hat{e}}
      \wedge \hat{R}^{\hat{a}\hat{b}}(\hat{\omega})
  \equiv
  \int \hat{\mathbf{L}}\,,
\end{equation}

\noindent
where $G_{N}^{(5)}$ is the 5-dimensional Newton constant. We assume KK
boundary conditions for the gravitational field: it admits a spatial Killing
vector $k$ with closed orbits with no fixed points. We partially break the
invariance under 5-dimensional general coordinate transformations (GCTs) by
using as 5\textsuperscript{th} coordinate, $\hat{x}^{4}$, that we will denote
by $z$, the coordinate adapted to $\hat{k}$. Thus $k=\partial_{\underline{z}}$
and all the components of the Vielbein are $z$-independent. Furthermore, since
$z$ parametrizes the orbits of $\hat{k}$, $z\sim z+2\pi \ell$ where $\ell$ is
some length scale. This length scale must be distinguished from the length of
the compact direction at a point $x$ of the 4-dimensional spacetime

\begin{equation}
  L(x)
  =
  2\pi R(x)
  =
  \int_{0}^{2\pi \ell} dz \sqrt{-\hat{g}_{\underline{z}\underline{z}}(x)}
  =
  2\pi \ell k(x)\,,
\end{equation}

\noindent
where $R(x)$ is the radius of the compact direction at the point $x$.
The value of the radius at infinity is customarily denoted by $R$ and it is
related to the value of the KK scalar at infinity $k_{\infty}$ by

\begin{equation}
  R
  =
  \ell k_{\infty}\,.
\end{equation}

Furthermore, we also partially break the invariance under 5-dimensional local
Lorentz transformations with the following choice of Vielbein\footnote{here
  $\imath_{\xi}$ indicates the interior product with the 4-dimensional vector
  $\xi$ and $\imath_{a}$ the interior product with $e_{a}$. Thus,
  $\imath_{a}A = e_{a}{}^{\mu}A_{\mu}$.  }

\begin{subequations}
  \label{eq:Vielbeinbasis}
  \begin{align}
    \hat{e}^{a}
    & =
      e^{a}\,, \hspace{1cm}
    &
      \hat{e}_{a}
    & =
      e_{a}-\imath_{a}A \partial_{\underline{z}}\,,
    \\
    & & & \nonumber \\
    \hat{e}^{z}
    & =
      k(dz+A)\,,
    &
      \hat{e}_{z}
    & =
      k^{-1}\partial_{\underline{z}}\,,
  \end{align}
\end{subequations}

\noindent
in which $e^{a},A$ and $k$ will be identified with the 4-dimensional,
KK-frame,\footnote{In this conformal frame, the Einstein--Hilbert term in the
  action carries an additional scalar factor $k$. A local conformal rescaling
  of the metric and a global rescaling of the KK vector will be necessary to
  eliminate this factor and obtain the Einstein-frame action with no additional
  constants in it.} Vielbein, KK vector and KK scalar. Therefore, the
5-dimensional metric and 4-dimensional KK-frame fields are related by

\begin{subequations}
  \begin{align}
    ds_{(5)}^{2}
   & =
    ds_{(4)}^{2} -k^{2}\left(dz+A\right)^{2}\,,
    \\
    & \nonumber \\
    ds_{(4)}^{2}
    & =
      g_{\mu\nu}dx^{\mu}dx^{\nu}\,.
  \end{align}
\end{subequations}

Under these assumptions and with these definitions, the action
Eq.~(\ref{eq:5dEHactionVielbein}) and after integration over the coordinate
$z$, can be rewritten in the form

\begin{equation}
  \label{eq:4dEHactionVielbein0}
  S[e,A,k]
  =
  \frac{2\pi \ell}{16\pi G_{N}^{(5)}}
  \int \left\{ k\left[ -\star(e^{a}\wedge e^{b})
      \wedge R_{ab} +\tfrac{1}{2}k^{2}F\wedge \star F \right]
    +d\left[2\star dk\right]\right\}\,.
\end{equation}

\noindent
Finally, the rescalings 

\begin{equation}
  \label{eq:rescalingsEinsteinframe}
  g_{\mu\nu}= \left(k/k_{\infty}\right)^{-1}g_{E\, \mu\nu}\,,
  \hspace{.5cm}
  e^{a}{}_{\mu} = \left(k/k_{\infty}\right)^{-1/2}e_{E}{}^{a}{}_{\mu}\,,
  \hspace{.5cm}
  A_{\mu} = k_{\infty}^{1/2}A_{E\, \mu}\,,
\end{equation}

\noindent
bring us to the Einstein-frame action\footnote{The constant (modulus)
  $k_{\infty}$ must be introduced in the rescaling in order to ensure that the
  normalization of the 5- and 4-dimensional metrics at spatial infinity are
  the same. Sometimes this conformal frame is referred to as the
  \textit{modified Einstein-frame} \cite{Maldacena:1996ky} to distinguish it
  from the one (standard in the earlier literature) in which this fact was not
  taken into account. In presence of matter fields, $k_{\infty}$ appears in
  the (modified) Einstein frame and one has to rescale some of the matter
  fields with it in order to make it disappear again. This is the origin of
  the definition of $A_{E}$. $k$ should not be rescaled if we do not want to
  modify its asymptotic value.}

\begin{equation}
  \label{eq:4dEHactionVielbeinEframe}
  \begin{aligned}
    S[e_{E},A_{E},k]
    & =
      \frac{1}{16\pi G_{N}^{(4)}}
      \int \left\{ -\star_{E}(e_{E}{}^{a}\wedge e_{E}{}^{b})
      \wedge R_{E\, ab}
      +\tfrac{3}{2}d\log{k}\wedge \star_{E} d\log{k}
       \right.
    \\
    & \\
    & \hspace{.5cm}
    \left.
      +\tfrac{1}{2}k^{3}F_{E}\wedge \star F_{E}
      +d\left[-\star_{E}d\log{k}\right]\right\}\,,
      \\
  \end{aligned}
\end{equation}

\noindent
where the 4-dimensional Newton constant is given by 

\begin{equation}
  \label{eq:4-5Newtonconstant}
  G_{N}^{(4)}
  =
  \frac{G_{N}^{(5)}}{2\pi R}\,.
\end{equation}

We can express the KK scalar in terms of an unconstrained\footnote{$k$ has to
  be positive definite.} scalar field $k = e^{\phi/\sqrt{3}}$ with
canonically-normalized kinetic term, but we will keep working with $k$ to
keep the discussion as simple as possible.

\subsection{The symmetries of 5-dimensional GR with KK boundary conditions}

In view of the results of Ref.~\cite{Gomez-Fayren:2024cpl} it is convenient to
revise the relation between the symmetries of the 5-dimensional theory and how
they are related to the those of the 4-dimensional one.

The conventional wisdom is that the only symmetries (local or global) of GR
are GCTs and, in our case, only those GCTs that preserve the KK boundary
conditions. In the Vielbein formulation we must also include local Lorentz
transformations that preserve those conditions and our choice
Eq.~(\ref{eq:Vielbeinbasis}). It turns out that GR with KK boundary conditions
also admits higher-form-type global symmetries which had traditionally been
incorrectly identified with a 1-parameter family of GCTs (see, for instance,
Refs.~\cite{Ortin:2015hya,Gomez-Fayren:2023wxk}).

Let us start with GCTs, which act on the 5-dimensional metric as

\begin{equation}
  \delta_{\hat{\xi}}\hat{g}_{\hat{\mu}\hat{\nu}}
  =
  -\mathcal{L}_{\hat{\xi}}\hat{g}_{\hat{\mu}\hat{\nu}}
  =
  -\left(\hat{\xi}^{\hat{\rho}}\partial_{\hat{\rho}}\hat{g}_{\hat{\mu}\hat{\nu}}
    +2\partial_{(\hat{\mu}}\hat{\xi}^{\hat{\rho}}\hat{g}_{\hat{\nu})\hat{\rho}}
  \right)\,.
\end{equation}

\noindent
In order to preserve the $z$-independence of the metric, all the components
except for $\hat{\xi}^{\underline{z}}$ must be $z$-independent. Furthermore,
the dependence $\hat{\xi}^{\underline{z}}$ on $z$ must be, at most,
linear. But vectors proportional to $z$ are not well defined because $z$ is
multivalued. For this reason, the vector field $z\partial_{\underline{z}}$,
which apparently generates constant rescalings of the 4-dimensional matter
fields which are, actually a global symmetry of the 4-dimensional action
Eq.~(\ref{eq:4dEHactionVielbeinEframe}), must be excluded. We will explain the
5-dimensional origin of that 4-dimensional symmetry in a moment.

Thus, taking into account the $z$-independence of the vector field
$\hat{\xi}$, we can write the action of the 5-dimensional GCTs that respect
the KK boundary conditions on the KK-frame 4-dimensional fields as

\begin{subequations}
  \begin{align}
    \delta_{\hat{\xi}} k
    & =
      -\hat{\xi}^{\rho}\partial_{\rho}k
      \,,
    \\
    & \nonumber \\
    \delta_{\hat{\xi}}A_{\mu}
    & =
      -\left(\hat{\xi}^{\rho}\partial_{\rho} A_{\mu}
      +\partial_{\mu}\hat{\xi}^{\rho} A_{\rho}\right)
      -\partial_{\mu}\hat{\xi}^{\underline{z}}\,,
    \\
    & \nonumber \\
    \delta_{\hat{\xi}}g_{\mu\nu}
    & =
      -\left(\hat{\xi}^{\rho}\partial_{\rho}g_{\mu\nu}
      +2\partial_{(\mu}\hat{\xi}^{\rho}g_{\nu)\rho}
      \right)\,,
  \end{align}
\end{subequations}

\noindent
which correspond to 4-dimensional GCTs generated by the 4-dimensional vector
field $\xi$ with components $\xi^{\mu}(x)=\hat{\xi}^{\mu}(x)$ and gauge
transformations of the KK vector fields $A$ generated by the 4-dimensional
gauge parameter

\begin{equation}
  \label{eq:chi}
  \hat{\xi}^{\underline{z}}(x)
  =
  -\chi(x)\,,
\end{equation}

\noindent
which act in the standard fashion: Lie derivative of the fields and

\begin{equation}
  \label{eq:deltachiA}
\delta_{\chi}A = d\chi\,,
\end{equation}

\noindent
respectively. Notice that these gauge transformations act on the compact
coordinate $z$:

\begin{equation}
  \label{eq:ztrans}
  \delta_{\chi}z
  =
  -\chi(x)\,,
  \,\,\,\,\,
  \Rightarrow
  \,\,\,\,\,
  \delta_{\chi}dz
  =
  -d\chi\,.
\end{equation}

The 5-dimensional local Lorentz group is broken by our choice of
(upper-triangular) Vielbein basis down to just the 4-dimensional one.

Let us now go back to 5-dimensional origin of the global symmetry of the
4-dimensional action Eq.~(\ref{eq:4dEHactionVielbeinEframe}) whose
transformations have the form

\begin{equation}
  \label{eq:4dimensionalrescalingsymmetry}
  \delta_{\epsilon}k
  =
  -\tfrac{2}{3}\epsilon k\,,
  \hspace{1cm}
  \delta_{\epsilon}A_{E}
  =
 \epsilon A_{E}\,.
\end{equation}

In Ref.~\cite{Gomez-Fayren:2024cpl} it was shown that the global
transformation

\begin{equation}
  \delta^{h}_{\epsilon}\hat{g}_{\hat{\mu}\hat{\nu}}
  =
  -2\epsilon \mathfrak{h}_{(\hat{\mu}}\hat{k}_{\hat{\nu})}\,,
\end{equation}

\noindent
where $\mathfrak{h}=\mathfrak{h}_{\hat{\mu}}d\hat{x}^{\hat{\mu}}$ is a
harmonic 1-form and $\hat{k}$ is the Killing vector of the theory rescales the
Einstein--Hilbert action with KK boundary conditions

\begin{equation}
  \delta^{h}_{\epsilon}\hat{S}
  =
  -\epsilon\,\imath_{k}\mathfrak{h}\,\hat{S}
  =
  -\epsilon \hat{S}\,
\end{equation}

\noindent
where we assumed the normalization $\imath_{\hat{k}}\mathfrak{h}=1$.  A
harmonic form $\mathfrak{h}$ typically exists in spacetimes with a compact
direction such as those satisfying KK boundary conditions. Locally, and up to
the total derivative of a $z$-independent function, $\mathfrak{h}$ can be
written as $dz$. The addition of a total derivative to $\mathfrak{h}$ is fully
equivalent to the transformation Eq.~(\ref{eq:ztrans}) and the gauge
transformation Eq.~(\ref{eq:deltachiA}). On the other hand, if $\mathfrak{h}$
is exact, then this transformation is equivalent to a GCT.

There is a second, independent, transformation that rescales the action

\begin{equation}
  \delta^{s}_{\epsilon} \hat{g}_{\hat{\mu}\hat{\nu}}
  =
  \tfrac{2}{3}\epsilon\hat{g}_{\hat{\mu}\hat{\nu}}\,,
  \,\,\,\,\,
  \Rightarrow
  \,\,\,\,\,
  \delta^{s}_{\epsilon}\hat{S}
  =
  +\epsilon \hat{S}\,,
\end{equation}

\noindent
and both transformations can be combined into a global symmetry of the
Einstein--Hilbert action with KK boundary conditions\footnote{This symmetry
  can be extended to include the coupling to matter fields \cite{kn:BCG-FMO}.}
$\delta_{\epsilon} \equiv \delta^{h}_{\epsilon}+\delta^{s}_{\epsilon}$ whose
effect on the 4-dimensional fields is precisely
Eq.~(\ref{eq:4dimensionalrescalingsymmetry}) \cite{Gomez-Fayren:2024cpl}.

In what follows we are going to need the transformations of the Vielbein and
spin connection under the above transformations. They are

\begin{subequations}
  \label{eq:Vielbeintrans}
  \begin{align}
   \delta^{h}_{\epsilon}\hat{e}^{\hat{a}}
    & =
      -\epsilon \imath_{\hat{k}} e^{\hat{a}}\mathfrak{h}\,, \hspace{1cm}
    &
   \delta^{h}_{\epsilon}\hat{\omega}^{\hat{a}\hat{b}}
    & =
      -\epsilon \hat{P}_{\hat{k}}{}^{\hat{a}\hat{b}}\mathfrak{h}\,,
      \hspace{1cm}
      &
        \delta^{h}_{\epsilon}\hat{R}^{\hat{a}\hat{b}}
    & =
      \epsilon \imath_{\hat{k}}\hat{R}^{\hat{a}\hat{b}}\wedge \mathfrak{h}\,,
    \\
    & & & \nonumber \\
   \delta^{s}_{\epsilon}\hat{e}^{\hat{a}}
    & =
      \tfrac{1}{3}\epsilon \hat{e}^{\hat{a}}
    &
   \delta^{s}_{\epsilon}\hat{\omega}^{\hat{a}\hat{b}}
    & =
      0\,,
      \hspace{1cm}
      &
        \delta^{s}_{\epsilon}\hat{R}^{\hat{a}\hat{b}}
    & =
      0\,,
  \end{align}
\end{subequations}

\noindent
where $\hat{P}_{\hat{k}}{}^{\hat{a}\hat{b}}$ is the \textit{Lorentz momentum
  map} or \textit{Killing bilinear}, defined by the \textit{momentum map
  equation}\footnote{We have used this equation together with the Palatini
  identity
  $\delta \hat{R}^{\hat{a}\hat{b}} = \mathcal{D}\delta\hat{\omega}^{\hat{a}\hat{b}}$
  to find $\delta^{h}_{\epsilon}\hat{R}^{\hat{a}\hat{b}}$ above.}

\begin{equation}
\imath_{\hat{k}}\hat{R}^{\hat{a}\hat{b}}
+\mathcal{D}\hat{P}_{\hat{k}}{}^{\hat{a}\hat{b}}
=
0\,,
\end{equation}

\noindent
which always admits the solution

\begin{equation}
  \hat{P}_{\hat{k}\, \hat{a}\hat{b}}
  =
  \mathcal{D}_{\hat{a}}\hat{k}_{\hat{b}}\,.
\end{equation}

In this language it is very easy to obtain the transformation of the
Einstein--Hilbert action Eq.~(\ref{eq:5dEHactionVielbein}). Ignoring the
normalization factor $(16\pi G_{N}^{(5)})$

\begin{equation}
  \begin{aligned}
    \delta^{h}_{\epsilon}\hat{S}
    & =
      \int \tfrac{1}{3!}\hat{\varepsilon}_{\hat{c}\hat{d}\hat{e}\hat{a}\hat{b}}\left\{      
 3\delta^{h}_{\epsilon}\hat{e}^{\hat{c}}\wedge \hat{e}^{\hat{d}} \wedge \hat{e}^{\hat{e}}
      \wedge \hat{R}^{\hat{a}\hat{b}}
      +\hat{e}^{\hat{c}}\wedge \hat{e}^{\hat{d}} \wedge \hat{e}^{\hat{e}}
      \wedge \delta^{h}_{\epsilon}\hat{R}^{\hat{a}\hat{b}}\right\}
    \\
    & \\
    & =
      -\epsilon\int \tfrac{1}{3!}\hat{\varepsilon}_{\hat{c}\hat{d}\hat{e}\hat{a}\hat{b}}\left\{      
 3\imath_{\hat{k}}e^{\hat{c}}\hat{e}^{\hat{d}} \wedge \hat{e}^{\hat{e}}
      \wedge \hat{R}^{\hat{a}\hat{b}}
      -\hat{e}^{\hat{c}}\wedge \hat{e}^{\hat{d}} \wedge \hat{e}^{\hat{e}}
      \wedge  \imath_{\hat{k}}\hat{R}^{\hat{a}\hat{b}}\right\}\wedge
      \mathfrak{h}
          \\
    & \\
    & =
      -\epsilon \int \imath_{\hat{k}}\hat{\mathbf{L}} \wedge
      \mathfrak{h}
      \\
    & \\
    & =
      -\epsilon \hat{S}\,.
  \end{aligned}
\end{equation}

On the other hand, we can use the generic variation of the action

\begin{equation}
  \delta \hat{S}
  =
  \int \left\{\hat{\mathbf{E}}_{\hat{a}}\wedge \delta\hat{e}^{\hat{a}}
  +d\hat{\mathbf{\Theta}}(\hat{e},\delta\hat{e}) \right\}\,,
\end{equation}

\noindent
where the Einstein equations and presymplectic potential are given by 

\begin{subequations}
  \begin{align}
    \hat{\mathbf{E}}_{\hat{a}}
    & =
      \imath_{\hat{a}}\hat{\star}(\hat{e}^{\hat{a}}\wedge\hat{e}^{\hat{b}})\wedge
      \hat{R}_{\hat{a}\hat{b}}\,,
    \\
    & \nonumber \\
    \mathbf{\Theta}(\hat{e},\delta\hat{e})
    & =
      -\hat{\star}(\hat{e}^{\hat{a}}\wedge\hat{e}^{\hat{b}})\wedge
      \delta \hat{\omega}_{\hat{a}\hat{b}}\,,
  \end{align}
\end{subequations}

\noindent
respectively, to find an alternative expression for $\delta^{h}_{\epsilon}S$
and arrive at the identity

\begin{equation}
-\epsilon\hat{\mathbf{L}}
=
\hat{\mathbf{E}}_{\hat{a}}\wedge \delta^{h}_{\epsilon}\hat{e}^{\hat{a}}
+d\hat{\mathbf{\Theta}}(\hat{e},\delta^{h}_{\epsilon}\hat{e}),
\end{equation}

\noindent
which, gives as an expression for the on-shell Lagrangian as a total
derivative\footnote{We use $\doteq$ for identities which hold on-shell.}

\begin{equation}
\hat{\mathbf{L}}
\doteq
  d\hat{\mathbf{J}}^{h}\,,
\end{equation}

\noindent
where we have defined the 4-form 

\begin{equation}
  \hat{\mathbf{J}}^{h}
  \equiv
  -\frac{1}{16\pi G_{N}^{(5)}}\left[\hat{\star}(\hat{e}^{\hat{a}}\wedge\hat{e}^{\hat{b}})
  \hat{P}_{\hat{k}\,\hat{a}\hat{b}}\right]  \wedge \mathfrak{h}\,.
\end{equation}

We may use this result as explained in Ref.~\cite{kn:CO} to determine the
generalized Komar charge of the theory. However, we observe that the 3-form
that multiplies $\mathfrak{h}$ is nothing but the Noether--Wald charge
associated to the Killing vector $\hat{k}$ of 5-dimensional GR.  In this case,
the Noether--Wald charge is nothing but (minus) the on-shell closed Komar
charge \cite{Komar:1958wp}

\begin{equation}
  \label{eq:Komarcharge5d}
  \hat{\mathbf{K}}[\hat{k}]
  =
  \frac{1}{16\pi G_{N}^{(5)}} \hat{\star}(\hat{e}^{\hat{a}}\wedge\hat{e}^{\hat{b}})
  \hat{P}_{\hat{k}\,\hat{a}\hat{b}}\,,
\end{equation}

\noindent
and we can write

\begin{equation}
  \label{eq:J=Kh}
  \hat{\mathbf{J}}^{h}
  =
  -\hat{\mathbf{K}}[\hat{k}]\wedge \mathfrak{h}\,.
\end{equation}

Thus, it follows that $\hat{\mathbf{J}}^{h}$ is actually closed (conserved)
and that the Lagrangian vanishes identically on-shell (a well-known fact in
pure GR).

We could have arrived at the same result using $\delta^{s}_{\epsilon}$ 
for which we find

\begin{equation}
\epsilon\hat{\mathbf{L}}
=
\hat{\mathbf{E}}_{\hat{a}}\wedge \delta^{s}_{\epsilon}\hat{e}^{\hat{a}}\,,
\end{equation}

\noindent
with no total derivative, owing to the invariance of the spin connection under
this rescaling $\delta^{s}_{\epsilon}\hat{\omega}^{\hat{a}\hat{b}}=0$.

Combining these two facts, we arrive at the conclusion that
$\hat{\mathbf{J}}^{h}$ is actually minus the Noether current associated to the
global symmetry
$\delta_{\epsilon} \equiv \delta^{h}_{\epsilon}+\delta^{s}_{\epsilon}$, which
explains its on-shell conservation.

\subsection{The 5-dimensional geometry of 4-dimensional stationary KK black
  holes}

According to the rigidity theorem \cite{Hawking:1971vc,Hawking:1973uf}, the
4-dimensional geometry of 4-dimensional, asymptotically flat, stationary black
holes is characterized by the existence of a Killing vector $m$ which is
timelike close to infinity and a spacelike Killing vector field $n$ that
generates rotations around a given axis. The Killing vector
$l\equiv m-\Omega_{\mathcal{H}}n$, where the constant $\Omega_{\mathcal{H}}$
is the angular velocity of the horizon, becomes null over the event horizon
$\mathcal{H}$

\begin{equation}
  l^{\mu}g_{\mu\nu}l^{\nu}
  \stackrel{\mathcal{H}}{=}
  0\,,
\end{equation}

\noindent
and $\mathcal{H}$ is, thus, identified as a Killing horizon.\footnote{Notice
  that the above equation takes the same form in the Einstein and the KK frame
  because conformal transformations preserve the lightcone.}

If this 4-dimensional black hole is a solution of the 4-dimensional KK theory
Eq.~(\ref{eq:4dEHactionVielbeinEframe}) that we have obtained by dimensional
reduction of GR with KK boundary conditions, it is also a solution of this
5-dimensional theory. As shown in \cite{Gomez-Fayren:2023wxk}, the
5-dimensional, asymptotically-KK solution is also stationary ($m$ is still a
Killing vector and it is timelike near infinity) and also has a Killing
horizon which is essentially the local product (a fibration) of the compact
dimension with the 4-dimensional horizon. The Killing vector that becomes null
over the 5-dimensional horizon (that we can call \textit{the uplift} of $l$)
is, actually \cite{Gomez-Fayren:2023wxk}

\begin{equation}
  \label{eq:lhatdef}
  \hat{l}
  \equiv
  l-\chi_{l}\hat{k}
  =
  m-\Omega_{\mathcal{H}}n -\chi_{l}\hat{k}\,.
\end{equation}

\noindent
According to the discussion in the previous section and Eq.~(\ref{eq:chi}),
the fact that $\hat{l}^{\underline{z}}=-\chi_{\hat{l}}$ means that the GCT
generated by the 4-dimensional Killing vector will leave invariant the
5-dimensional metric if it is supplemented by a \textit{compensating} or
\textit{induced gauge transformation} with parameter $\chi_{l}$, whose value
we can determine by solving the 5-dimensional Killing vector equation for
$\hat{l}$, $\mathcal{L}_{\hat{l}}\hat{g}_{\hat{\mu}\hat{\nu}}=0$, assuming
that $\hat{g}_{\hat{\mu}\hat{\nu}}$ satisfies the KK boundary conditions and
is $z$-independent. We find three independent conditions:

\begin{subequations}
  \begin{align}
    \mathcal{L}_{l}k
    & =
      0\,,
    \\
    & \nonumber \\
    \label{eq:secondcondition}
    \mathcal{L}_{l}A-d\chi_{l}
    & =
      0\,,
    \\
    & \nonumber \\
    \mathcal{L}_{l}g_{\mu\nu}
    & =
      0\,,
  \end{align}
\end{subequations}

The second condition Eq.~(\ref{eq:secondcondition}), which confirms the
interpretation of $\chi_{l}$ as an induced gauge transformation, can be
rewritten in the form of the \textit{momentum map equation}

\begin{equation}
  \label{eq:Pldef}
  \imath_{l}F +dP_{l}
  =
  0\,,
  \hspace{1cm}
  P_{l} = -\chi_{l}+\imath_{l}A\,.
\end{equation}

The symmetry condition $\mathcal{L}_{l}F=0$, together with the Bianchi
identity $dF$ guarantee the local existence of the \textit{momentum map}
$P_{l}$, which is determined by the above equation up to an additive constant.

Thus, 

\begin{equation}
  \label{eq:chil}
  \chi_{l}
  =
  \imath_{l}A-P_{l}\,.  
\end{equation}

Finally, the condition that $\mathcal{H}$ must be a Killing horizon of $\hat{l}$

\begin{equation}
  \hat{l}^{\hat{\mu}}\hat{g}_{\hat{\mu}\hat{\nu}}\hat{l}^{\hat{\nu}}
  =
  l^{\mu}g_{\mu\nu}l^{\nu}
 -k^{2}(\imath_{l}A-\chi_{l})^{2}
  \stackrel{\mathcal{H}}{=}
  0\,,
\end{equation}

\noindent
together with Eq.~(\ref{eq:chil}) imply that we must choose the additive
constant in $P_{l}$ so that

\begin{equation}
  P_{l}
    \stackrel{\mathcal{H}}{=}
  0\,.
\end{equation}

\section{The 5-dimensional generalized Komar charge}

Linear combinations of conserved charges with constant coefficients are
conserved. In our case this means that, as discussed in Ref.~\cite{kn:CO}, if
there is another conserved 3-form charge, we can add it to the standard Komar
charge Eq.~(\ref{eq:Komarcharge5d}) with an arbitrary coefficient that we must
determine using some physical criterion.

There are no other 3-form charges in 5-dimensional pure GR with
asymptotically-flat boundary conditions, but there is one with KK boundary
conditions, that we can derive from the Noether 4-form charge associated to
the higher-form symmetry identified in Ref.~\cite{Gomez-Fayren:2024cpl}.

As a general rule, given a $(d-1)$-form current $\mathbf{J}$ which is
conserved when evaluated over a solution of the theory with a spacetime
symmetry generated by the vector $p$ (\textit{i.e.}~$\delta_{p}$ annihilates
all the fields of the solution and, therefore, the current), it is always
possible to derive from it a $(d-2)$-form charge $\mathbf{Q}_{p}$ which is
also conserved under the same assumptions \cite{Ballesteros:2023iqb}.

In the case at hands, we are interested in stationary KK black holes which, on
top of $\hat{k}$, admit Killing vector $\hat{l}$ given in
Eq.~(\ref{eq:lhatdef}) and we can derive a 3-form charge
$\hat{\mathbf{Q}}^{h}_{\hat{l}}$ from the 4-form $\mathbf{J}^{h}$ using its
on-shell conservation and the assumption $\delta_{\hat{l}}\mathbf{J}^{h}=0$.

How does $\delta_{\hat{l}}$ act on $\mathbf{J}^{h}$? Let us first consider
how $-\mathcal{L}_{\hat{l}}$ acts on $\mathfrak{h}$:

\begin{equation}
  -\mathcal{L}_{\hat{l}}\mathfrak{h}
  =
  -d \imath_{\hat{l}}\mathfrak{h}
  =
  d\chi_{l}\,,
\end{equation}

\noindent
which, upon use of Eq.~(\ref{eq:ztrans}), we can rewrite in the form

\begin{equation}
  \delta_{\hat{l}}\mathfrak{h}
  =
  -\mathcal{L}_{\hat{l}}\mathfrak{h} +\delta_{\chi_{l}}\mathfrak{h}
  =
  0\,.
\end{equation}

On the other hand, $\delta_{\hat{l}}$ acts on $\hat{\mathbf{K}}[\hat{k}]$ as
minus the Lie derivative and, furthermore, since $\hat{l}$ is a Killing vector
and $\delta_{\hat{l}}$ annihilates the Vielbein, metric etc.,
$\delta_{\hat{l}}\hat{\mathbf{K}}[\hat{k}]=
\hat{\mathbf{K}}[\delta_{\hat{l}}\hat{k}]=0$ because
$\delta_{\hat{l}}\hat{k} = -[\hat{l},\hat{k}]=0$.

Then,

\begin{equation}
  0
  =
    \delta_{\hat{l}}\hat{\mathbf{J}}^{h}
 =
 -(d\imath_{\hat{l}} +\imath_{\hat{l}}d)\hat{\mathbf{J}}^{h}
-\hat{\mathbf{K}}[\hat{k}]\wedge\delta_{\chi_{\hat{l}}}\mathfrak{h}
 \doteq
 -d\left\{\imath_{\hat{l}}\hat{\mathbf{J}}^{h}
   +\chi_{\hat{l}}\hat{\mathbf{K}}[\hat{k}]\right\}\,,
\end{equation}

\noindent
and we can define the 3-form charge

\begin{equation}
  \label{eq:Qh}
  \hat{\mathbf{Q}}^{h}_{\hat{l}}
  \equiv
  \imath_{\hat{l}} \hat{\mathbf{J}}^{h}+\chi_{\hat{l}}\hat{\mathbf{K}}[\hat{k}]\,,  
\end{equation}

\noindent
which is conserved (closed) when the hypotheses are satisfied: on-shell for
solutions admitting the Killing vector $\hat{l}$. Since in
Ref.~\cite{Gomez-Fayren:2024cpl} it was shown that the 5-dimensional 4-form
current $\hat{\mathbf{J}}^{h}$ is related to the 4-dimensional 3-form current
associated to the global symmetry
Eq.~(\ref{eq:4dimensionalrescalingsymmetry}), we expect 3-form charge
Eq.~(\ref{eq:Qh}) to be related to the 2-form charge that one can derive from
the 4-dimensional current. The integral of this charge at infinity gives the
scalar charge of the black hole \cite{Pacilio:2018gom,Ballesteros:2023iqb}.

There are no more conserved charges in the theory under consideration and,
therefore, we are led to consider the 1-parameter family of generalized Komar
charges

\begin{equation}
  \label{eq:Kalpha}
  \begin{aligned}
    \hat{\mathbf{K}}_{\alpha}[\hat{l}]
    & \equiv
      \frac{1}{16\pi G_{N}^{(5)}}
      \left\{\hat{\mathbf{K}}[\hat{l}]
      -\alpha\imath_{\hat{l}}\left[ \hat{\mathbf{K}}[\hat{k}]
      \wedge \mathfrak{h}\right] +\alpha \chi_{\hat{l}} \hat{\mathbf{K}}[\hat{k}]
      \right\}
    \\
    & \\
    & =
      \frac{1}{16\pi G_{N}^{(5)}}
      \left\{ \hat{\mathbf{K}}[\hat{l}]
      -\alpha\imath_{\hat{l}}\hat{\mathbf{K}}[\hat{k}]
      \wedge \mathfrak{h}
      \right\}
  \end{aligned}
\end{equation}

\noindent
since, by assumption, $\imath_{\hat{k}}\mathfrak{h}=1$ and, therefore,
$\imath_{\hat{l}}\mathfrak{h}=-\chi_{\hat{l}}$.  In view of the previous
comments, we expect that its integral at infinity will remove the contribution
of the scalar charge mentioned in the Introduction (see
Eq.~(\ref{eq:scalarchargeplusmass})) for some value of $\alpha$. It is worth
stressing that, as explained in the introduction, $\alpha$ has no influence on
the Smarr formula.

In order to determine the value of $\alpha$ we have to study the pullback of
$\mathbf{K}_{\alpha}[\hat{l}]$ over hypersurfaces that include the compact
direction. The only components of $\mathbf{K}_{\alpha}[\hat{l}]$ whose
pullback does not identically vanish are those that contain $\mathfrak{h}$ as
factor.

For any 5-dimensional Killing vector $\hat{p}$, the standard Komar charge
Eq.~(\ref{eq:Komarcharge5d}) has the following decomposition:

\begin{equation}
  \hat{\mathbf{K}}[\hat{p}]
  =
  \frac{1}{16\pi G_{N}^{(5)}}
  \left[-k\star (e^{a}\wedge e^{b})\hat{P}_{\hat{p}\, ab}\right]\wedge \mathfrak{h}
  +2\star e^{a}\hat{P}_{\hat{p}\, az}
  -\left[k\star (e^{a}\wedge e^{b})\hat{P}_{\hat{p}\, ab}\right]\wedge A\,,
\end{equation}

\noindent
and, therefore, the terms of $\mathbf{K}_{\alpha}[\hat{l}]$ that contribute to
the pullback are

\begin{equation}
  \label{eq:Kalpha2}
  \begin{aligned}
  \hat{\mathbf{K}}_{\alpha}[\hat{l}]_{\mathfrak{h}}
  & =
    \frac{1}{16\pi G_{N}^{(5)}}
    \left\{-k\star (e^{a}\wedge e^{b})\left(\hat{P}_{\hat{l}\, ab}
    +\alpha\chi_{l}\hat{P}_{\hat{k}\, ab}\right)
    \right.
    \\
    & \\
    & \hspace{.5cm}
      \left.
    -\alpha\imath_{\hat{l}}\left[2\star e^{a}\hat{P}_{\hat{k}\, az}
  -\left[k\star (e^{a}\wedge e^{b})\hat{P}_{\hat{k}\, ab}\right]\wedge A\right]
  \right\}\wedge \mathfrak{h}\,.
  \end{aligned}
\end{equation}

Using the relation between the 5-dimensional Killing vector $\hat{l}$, the
4-dimensional one $l$ and $\hat{k}$ Eq.~(\ref{eq:lhatdef}) and the expression
Eq.~(\ref{eq:chil}) for the parameter of the compensating gauge
transformation, we find

\begin{equation}
  \begin{aligned}
  \hat{P}_{\hat{k}\, ab}
  & =
  -\tfrac{1}{2}k^{2}F_{ab}\,,
    \hspace{1cm}
    &
  \hat{P}_{\hat{k}\, az}
  & =
    -\partial_{a}k\,,
    \\
    & & & \\
  \hat{P}_{\hat{l}\, ab}
  & =
  P_{l\, ab}-\tfrac{1}{2}P_{l}k^{2}F_{ab}\,,
    \hspace{1cm}
    &
  \hat{P}_{\hat{l}\, az}
  & =
    -\tfrac{1}{2}k^{-1}\partial_{a}\left(k^{2} P_{l}\right)\,,
  \end{aligned}
\end{equation}

\noindent
and, replacing these values in Eq.~(\ref{eq:Kalpha2}) we get

\begin{equation}
  \label{eq:Kalpha3}
  \begin{aligned}
  \hat{\mathbf{K}}_{\alpha}[\hat{l}]_{\mathfrak{h}}
    & =
      \frac{1}{16\pi G_{N}^{(5)}}\left\{-k\star d\mathbf{l}
          -2\alpha\star( dk\wedge \mathbf{l})
    +(1-\alpha)P_{l}k^{3}\star F
        -\alpha \tilde{P}_{l} F
        \right\}\wedge \mathfrak{h}
    \\
    & \\
    & \hspace{.5cm}
        +d\left(
        \alpha \tilde{P}_{l} A \wedge \mathfrak{h} \right)\,,
  \end{aligned}
\end{equation}

\noindent
where we have denoted by $\mathbf{l}$ the 1-form dual to the 4-dimensional
Killing vector $l$ and we have defined the \textit{dual (magnetic) momentum
  map} $\tilde{P}_{l}$.\footnote{Since
  \begin{equation}
    d\left(k^{3}\star F\right) \doteq 0\,,
    \,\,\,\,\,
    \text{and}
    \,\,\,\,\,
    \mathcal{L}_{l}\left(k^{3}\star F\right) =0\,,
  \end{equation}
  by assumption, $\imath_{l}\left(k^{3}\star F\right)$ is closed on-shell, and
  there exists a function $\tilde{P}_{l}$ satisfying the \textit{dual momentum
    map equation}
  \begin{equation}
    \imath_{l}\left(k^{3}\star F\right)
    \doteq
    -d\tilde{P}_{l}\,. 
  \end{equation}
} The components of this 1-form are computed lowering the
vector index with the KK-frame metric. Thus, according to
Eq.~(\ref{eq:rescalingsEinsteinframe}),

\begin{equation}
  \mathbf{l}
  =
  l^{\mu}g_{\mu\nu}dx^{\nu}
  =
  \left(k/k_{\infty}\right)^{-1}l^{\mu}g_{E\, \mu\nu}dx^{\nu}
  =
  \left(k/k_{\infty}\right)^{-1} \mathbf{l}_{E}\,.
\end{equation}

\noindent
Defining

\begin{subequations}
  \begin{align}
    P_{l}
    & \equiv
      k^{1/2}_{\infty}P_{E\, l}\,,
      \hspace{1cm}
    & 
      \imath_{l}F_{E}
    & =
      -dP_{E\, l}\,,
    \\
    & & & \nonumber \\
    \tilde{P}_{l}
    & \equiv
      k^{1/2}_{\infty}\tilde{P}_{E\, l}\,,
      \hspace{1cm}
    & 
      \imath_{l}\left(k^{3}\star_{E} F_{E}\right)
    & =
      -dP_{E\, l}\,,
  \end{align}
\end{subequations}

\noindent
and, discarding the total derivative, we arrive at\footnote{Notice that, for
  2-forms $G$, $\star G=\star_{E} G$.}

\begin{equation}
  \label{eq:Kalpha4}
  \hat{\mathbf{K}}_{\alpha}[\hat{l}]_{\mathfrak{h}}
    =
    \frac{k_{\infty}}{16\pi G_{N}^{(5)}}\left\{-\star_{E} d\mathbf{l}_{E}
      +(1-2\alpha)\star_{E} \left(d\log{k}\wedge\mathbf{l}_{E}\right)
    +(1-\alpha)P_{E\, l}k^{3}\star_{E} F_{E}
        -\alpha \tilde{P}_{E\, l} F_{E}\right\}
      \wedge \mathfrak{h}\,.
\end{equation}

The first term in $\hat{\mathbf{K}}_{\alpha}[\hat{l}]_{\mathfrak{h}}$ is the
standard 4-dimensional Komar charge associated to the 4-dimensional Killing
vector $l$. That charge is not closed and two additional terms similar to the
third and the fourth but with different coefficients, have to be added to
construct the on-shell closed generalized Komar charge of the KK theory
Eq.~(\ref{eq:4dEHactionVielbeinEframe}). The second term, on the other hand,
gives an unwanted additional contribution which is proportional to the scalar
charge of the KK scalar $k$. The same value of $\alpha$ that kills that term
($\alpha=1/2$) give the right coefficients of the third and fourth terms and,
using the relation between the 4- and 5-dimensional Newton constants
Eq.~(\ref{eq:4-5Newtonconstant}), we finally arrive at

\begin{equation}
  \label{eq:Kalpha5}
  \begin{aligned}
  \hat{\mathbf{K}}_{1/2}[\hat{l}]_{\mathfrak{h}}
    & =
      \frac{k_{\infty}}{16\pi G_{N}^{(5)}}\left\{
      -\star_{E} d\mathbf{l}_{E}
    +\tfrac{1}{2}P_{E\, l}k^{3}\star_{E} F_{E}
        -\tfrac{1}{2}\tilde{P}_{E\, l} F_{E}\right\}
      \wedge \mathfrak{h}
    \\
    & \\
    & =
 \mathbf{K}[l]\wedge \frac{\mathfrak{h}}{2\pi \ell}\,,
  \end{aligned}
\end{equation}

\noindent
which, integrated over the compact direction gives the 4-dimensional
generalized Komar 2-form charge.

Given the normalization of the momentum map $P_{E\, l}$ (we had to impose that
it vanishes over the horizon), the integral at infinity will include a
contribution $\sim \Phi_{\infty}Q$ where $\Phi_{\infty}$ is the value of the
corotating electrostatic potential at infinity.\footnote{We are free to impose
  that $\tilde{P}_{E\, l}$ vanishes at infinity or, at least, that the
  integral of $\tilde{P}_{E\, l}F_{E}$ does \cite{Zatti:2024vbz}.}  In
general, that value will not vanish. However, we are free to add on-shell
closed 2-form charges like
$-\Phi_{\infty} k^{3}\star F_{E}\wedge \mathfrak{h}$ to remove that
contribution. This can be done directly in 5 dimensions, adding a term
$\sim \Phi_{\infty} \hat{\mathbf{K}}[\hat{k}]$. In practice, this is
equivalent to replacing $\hat{l}$ by
$\hat{l}' \equiv \hat{l}-\Phi_{\infty}\hat{k}$ which has the same form as
$\hat{l}$ but with $P_{l}$ now vanishing at infinity instead of the horizon.

\section{Discussion}

In this paper we have shown how to construct a generalized Komar charge in
5-dimensional pure gravity with KK boundary conditions whose integral at
spatial infinity for the Killing vector that generates time translations gives
the mass, understood as the 4-dimensional mass, removing the contribution of
the 4-dimensional KK scalar charge.

The construction is based on the freedom that one has to construct conserved
charges as linear combinations of other conserved charges with constant
coefficients and to the existence of a higher-form-type global symmetry in GR
with KK boundary conditions \cite{Gomez-Fayren:2024cpl} out of which we can
derive conserved charges in symmetric solutions using the recipe of
\cite{Gomez-Fayren:2023wxk}.

Since this higher-form-type global symmetry can be extended to the coupling
with matter, it is interesting to see how the results of this paper can be
generalized to more complex situations. Work in this direction is well
underway \cite{kn:BCG-FMO}.

\section*{Acknowledgments}

The work of GB, CG-F, TO and JLVC has been supported in part by the MCI, AEI,
FEDER (UE) grants PID2021-125700NB-C21 (``Gravity, Supergravity and
Superstrings'' (GRASS)) and IFT Centro de Excelencia Severo Ochoa
CEX2020-001007-S.  The work of PM has been supported by the MCI, AEI, FEDER
(UE) grant PID2021-123021NB-I00.  The work of GB has been supported by the
fellowship CEX2020-001007-S-20-5. The work of CG-F was supported by the MU
grant FPU21/02222. The work of JLVC has been supported by the CSIC JAE-INTRO
grant JAEINT-24-02806. TO wishes to thank M.M.~Fern\'andez for her permanent
support.

\appendix


\end{document}